\documentclass[10pt, technote, onecolumn, draftcls]{IEEEtran}
\usepackage{amssymb} 
\usepackage{verbatim}
\usepackage{amsmath}
\usepackage{caption}
\usepackage{subfig} 
\usepackage[dvipsnames]{xcolor}
\usepackage{url}
\usepackage[compress]{cite}
\usepackage{tikz}
\usetikzlibrary{shapes}
\usepackage{soul}
\usepackage[pdftex, bookmarks, bookmarksopen, CJKbookmarks]{hyperref}
\usepackage{algorithm}
\usepackage{algorithmic}
\usepackage{bbm}

\newcommand{\dongningstyle}{true}
\newcommand{\fordiscussion}{false}
\newcommand{\MII}{\mathcal{I}}
\newcommand{\ENTH}{\mathcal{H}}
\usepackage{ifthen}
\usepackage{pgffor}

\ifthenelse{\equal{\dongningstyle}{true}}{
        \newcommand{\RVEC}[1]{\boldsymbol{\uppercase{#1}}}
        \newcommand{\RMAT}[1]{\underline{\boldsymbol{\uppercase{#1}}}}
        \newcommand{\RSCA}[1]{\uppercase{#1}}
        \newcommand{\VEC}[1]{\boldsymbol{\lowercase{#1}}}
        \newcommand{\MAT}[1]{\boldsymbol{\underline{\lowercase{#1}}}}
}{
        \newcommand{\RVEC}[1]{\boldsymbol{#1}}
        \newcommand{\RMAT}[1]{\boldsymbol{\mathrm{\uppercase{#1}}}}
        \newcommand{\RSCA}[1]{\uppercase{#1}}
        \newcommand{\VEC}[1]{\boldsymbol{\lowercase{#1}}}
        \newcommand{\MAT}[1]{\boldsymbol{\underline{\lowercase{#1}}}}
}

\ifthenelse{\equal{\fordiscussion}{true}}{
        \newcommand{\NONUM}{ }
}{
        \newcommand{\NONUM}{\nonumber}
}

\foreach \x in {A,...,Z}{%
         \expandafter\xdef\csname rv\x\endcsname{\noexpand\RVEC{\x}}
         \expandafter\xdef\csname rm\x\endcsname{\noexpand\RMAT{\x}}
         \expandafter\xdef\csname r\x\endcsname{\noexpand\RSCA{\x}}
         \expandafter\xdef\csname vv\x\endcsname{\noexpand\VEC{\x}}
         \expandafter\xdef\csname mm\x\endcsname{\noexpand\MAT{\x}}
}

\foreach \x in {a,...,z}{%
         \expandafter\xdef\csname rv\x\endcsname{\noexpand\RVEC{\x}}
         \expandafter\xdef\csname rm\x\endcsname{\noexpand\RMAT{\x}}
         \expandafter\xdef\csname vv\x\endcsname{\noexpand\VEC{\x}}
         \expandafter\xdef\csname mm\x\endcsname{\noexpand\MAT{\x}}
}

\newtheorem{theorem}{Theorem}
\newtheorem{remark}{Remark}

\newtheorem{proposition}{Proposition}
\newtheorem{lemma}{Lemma}
\newtheorem{conjecture}{Conjecture}

\newcommand{\MEXP}[1]{\mathbb{E}\left[ #1 \right]}
\newcommand{\MEXPP}{\mathbb{E}}
\newcommand{\Prob}[1]{\mathsf{P}\left( #1 \right)}

\newcommand{\Probb}{\mathsf{P}}

\newcommand{\MI}[1]{\ensuremath{\MII \left( #1 \right)}}
\newcommand{\CMI}[2]{\ensuremath{\MII \left(\left. #1 \right| #2 \right)}}
\newcommand{\CMIR}[2]{\ensuremath{\MII \left( #1 \left| #2 \right. \right)}}

\newcommand{\CENT}[2]{\ensuremath{\ENTH\left( \left. #1 \right| #2 \right) }}
\newcommand{\CENTR}[2]{\ensuremath{\ENTH\left( #1 \left| #2 \right. \right) }}

\newcommand{\IND}[1]{\mathbbm{1}_{\left( #1 \right)}}

\newcommand{\uT}{\mathsf{T}}

\newcommand{\IE}{{\em i.e.}}

\newcommand{\ETAL}{{\em et al.}}

\newcommand{\SEQ}[1]{\ensuremath{( #1 )}}
\newcommand{\SEQN}[1]{\ensuremath{(#1 )^n}}

\newcommand{\rvWt}{\widetilde{\rvW}}
\newcommand{\rvWb}{\overline{\rvW}}

\newcommand{\rNt}{\widetilde{\rN}}
\newcommand{\rLt}{\widetilde{\rL}}
\newcommand{\rMt}{\widetilde{\rM}}
\newcommand{\sLt}{\tilde{l}}
\newcommand{\sMt}{\widetilde{m}}
\newcommand{\sNt}{\tilde{n}}
\newcommand{\Ber}[1]{\mathrm{Ber}\left(#1\right)} 
\newcommand{\REGION}{\mathcal{R}}
\newcommand{\rlambda}{\RSCA{\Lambda}}
\newcommand{\BOUND}{\mathsf{B}}

\title{On Layered Erasure Interference Channels without CSI at Transmitters}
\author{
\IEEEauthorblockN{Yan Zhu}
\IEEEauthorblockA{Aerohive Network Inc. \\330 Gibraltar Dr, Sunnyvale, CA 94089 \\ Email: zhuyan981087@gmail.com}}
\begin{document}
\maketitle
\begin{abstract}
This paper studies a layered erasure model for two-user interference channels, which can be viewed as a simplified version
of Gaussian fading interference channel. It is assumed that channel state information~(CSI) is only available at receivers but not at
transmitters.  Under such assumption, an outer bound is derived for the capacity region of such interference channel. The
new outer bound is tight in many circumstances. For the remaining open cases, the outer bound extends previous results in~\cite{Vahid2015IC}. 
\end{abstract}
\section{Introduction}
Since the breakthrough of Gaussian interference channel by Etkin, Tse and Wang~\cite{Etkin08IFC}, the study of
interference networking has made plenty of progress\cite{Gamal2011network}. However, most works still focus on situations where
channel state information~(CSI) is static and known to both transmitters~(CSIT) and receivers~(CSIR). These results
usually valid under situations where CSI variates slowly and systems have efficient sounding and feedback mechanisms
to update global CSI timely. For communications experiencing fast channel fading, we usually do not have such luxury to have
(accurate and timely) CSI at transmitters.

In this paper, we investigate a layered erasure model of two-user interference channel~(IC), which shares the same
spirit as deterministic model used in~\cite{Bresler08IFC}, except that the transmit binary vectors are erased
randomly. We assume that the erasure levels (which model the fading states or CSI) are known at the receivers but not at
transmitters. In particular, we derive an outer bound for general
two-user layered erasure interference channels with no CSI at transmitters. The obtained new bound is generally tight for
many important cases but whether it is tight for all situations is still open in this paper. Comparing with previous
results in~\cite{Vahid2015IC}, this paper can be viewed as a fully extension to multi-layer situations. 

The remaining paper is organized as following. In next section, the channel model is formally described and some
notation is introduced to assist the further discussion. In Section~\ref{sec:mainresult}, our main finding is presented and followed by several remarks to
clarify the new outer bound. We continue our discussion by investigating some non-trivial
situations in Section~\ref{sec:LEC}.  Limited by the space, we only present a brief proof of the outer bound in
Section~\ref{sec:proof}. Finally, we conclude this paper in
Section~\ref{sec:conclude}.

\section{Model and Notations}\label{sec:intro}
\begin{figure}[h]
  \centering
  \includegraphics[width=.3\textwidth]{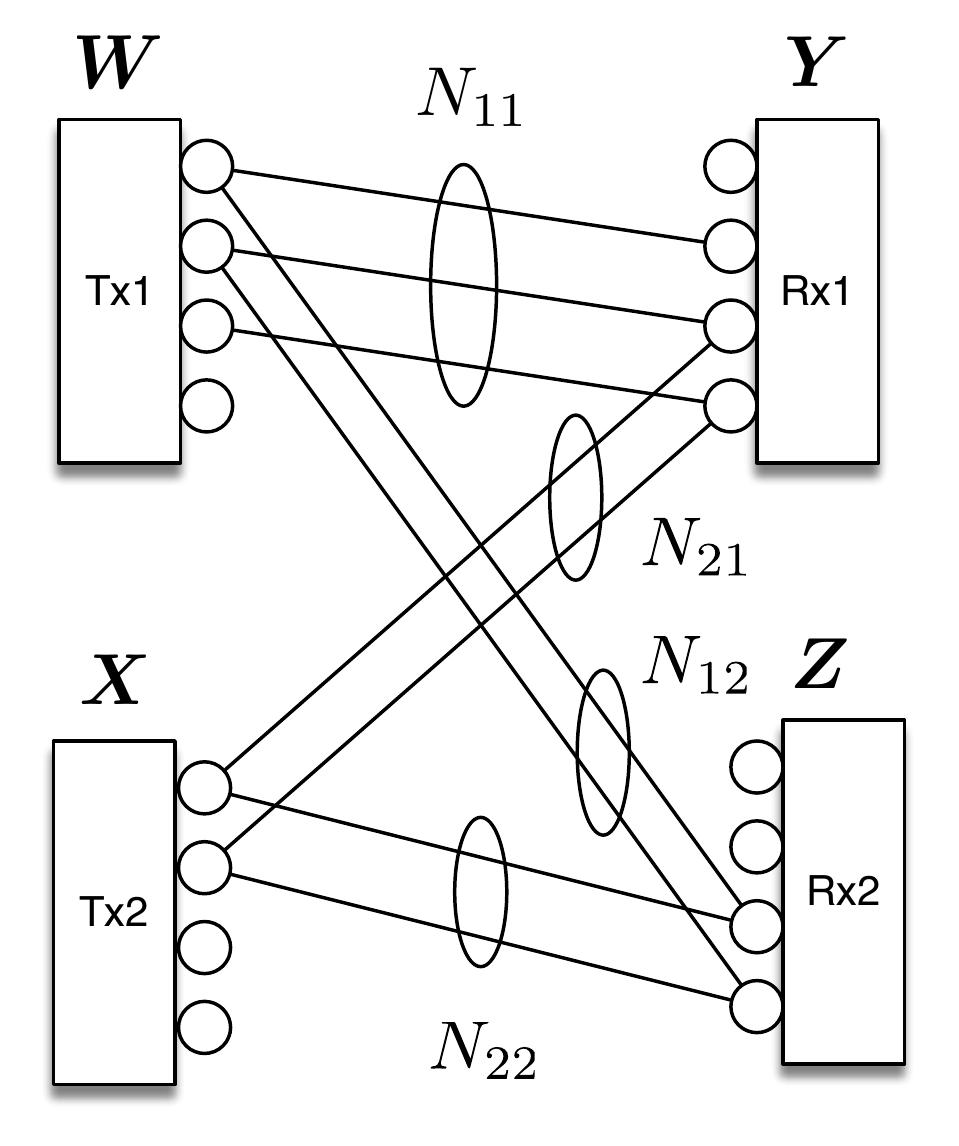}
  \caption{Layered Erasure Interference Channel Model. }
   \label{fig:model}
\end{figure}
Consider a layered erasure interference channel as shown in Fig.~\ref{fig:model}. At each time $i$,
transmitters~1 and~2 emit binary-vector signals
$\rvW[i]$ and $\rvX[i]$, respectively, which take value in $\mathbb{F}^q_2$. Only a certain 
top portion of each vector signal reaches the two receivers, and the remaining part is erased randomly. Mathematically, let $\mmS$
denote a $q\times q$ matrix whose elements are all $0$ except that
$\mmS_{k+1,k}=1$ for $k=1,\dots, q-1$. It is easy to see that $\mmS
[x_1,x_2,\dots,x_q]^\uT = [0, x_1,\dots,x_{q-1}]^\uT$, and $\mmS^{q-n} [x_1,x_2,\dots,x_q]^\uT = [0, \dots, 0, x_1,
\dots, x_n]^\uT$, which is equivalent to a zero-padding downward shift of the vector so that only its first $n$ elements are
left. With these notations, at each time $i$, the two received signals, $\rvY[i]$ and $\rvZ[i]$, can be written as,  
\begin{subequations}\label{eq:sys}
  \begin{align}
    \rvY[i] & = \mms^{q-\rN_{11}[i]} \rvW[i] \oplus \mms^{q-\rN_{21}[i]} \rvX[i] \label{eq:sys1}\\
    \rvZ[i] & = \mms^{q-\rN_{12}[i]} \rvW[i] \oplus \mms^{q-\rN_{22}[i]} \rvX[i], \label{eq:sys2}
  \end{align}
\end{subequations}
respectively, where for $t,r= 1,2$, each integer random process $\{\rN_{tr}[i]\}$ models the channel fading process from
transmitter $t$ to receiver $r$. We assume that the four fading processes are independent of each other and each of them is
independent and identically distributed~(i.i.d) over time (so that the channel is memoryless). In this paper, we study
situations where receiver~1 knows realization of $(\rN_{11}[i], \rN_{21}[i])$ and receiver~2 knows realization of
$(\rN_{22}[i], \rN_{12}[i])$ at each time $i$, but no channel state information~(CSI) is available at both transmitters
except for the statistical law of those fading processes. 

For remaining discussion, we need following notations. Suppose $\rvX \in \mathbb{F}^q_2$ is an arbitrary random vector.  We use $\rX_j$ to
denote its $j$-th element and $\rvX_j^k$ to denote its sub-vector $[\rX_j, \dots, \rX_k]^\uT$. For the special case
where subscript $j=1$, we often use $\rvX^k$ instead of $\rvX^k_1$. We often use lower-case letters to denote realizations
of their corresponding random vectors or random variables. For example, $\vvx = [x_1, x_2, \dots, x_q]^\uT$ should be interpreted as a particular
realization of the
random vector $\rvX$. For any sequence of random vectors $\{\rvX[i]\}$, let $(\rvX)_{i_1}^{i_2}$ denote the subsequence
$\rvX[i_1], \dots, \rvX[i_2]$. Consequently, $(\rvX_j^k)_{i_1}^{i_2}$ denotes the subsequence
$\rvX_j^k[i_1], \rvX_j^k[i_1+1], \dots, \rvX_j^k[i_2]$. In summary, the indices outside the parentheses always refer to
time while inside ones refer to element(s) of the corresponding vectors. Binary addition ($\oplus$) between two vectors with
different lengths is aligned at the least significant bits: if $n_1\geq n_2$, then define
$\rvX^{n_1} \oplus \rvW^{n_2}=[\rX_1, \dots, \rX_{n_1-n_2} , \rX_{n_1-n_2+1}\oplus \rW_1, \dots, \rX_{n_1} \oplus
\rW_{n_2}]^\uT$.
Since we only consider memoryless channels in this paper, we often surpress the time index
$i$ to ease the notations. For example, $\Prob{\rN_{11} \geq l}$ is equivalent to $\Prob{\rN_{11}[i] \geq l}$, distribution of
$\rN_{11}[i]$ is referred as distribution of $\rN_{11}$, \ETAL With the convention introduced above, 
channel model~\eqref{eq:sys} can be rewritten as
\begin{subequations}\label{eq:sys:sh}
  \begin{align}
    \rvY & = \rvW^{\rN_{11}} \oplus \rvX^{\rN_{21}} \label{eq:sys1:sh}\\
    \rvZ & = \rvW^{\rN_{12}} \oplus \rvX^{\rN_{22}}. \label{eq:sys2:sh}
  \end{align}
\end{subequations}

\section{Main Results}\label{sec:mainresult}

Our main findings are summarized in Theorem~\ref{thm:1}, which needs a few of more definitions. Define following three
regions within first orthant of $\mathbb{R}^2$ as
\begin{subequations} \label{eq:bounds}
  \begin{align}
    \REGION_{1a}  & := \bigcap_{\omega \in [0, 1]} \left\{ \left. (R_1, R_2) \in \mathbb{R}^2_{\geq 0} \right| R_1 + \omega R_2 \leq \BOUND_{1a}(\omega)
                        \right\} \label{eq:ZIC-bound} \\
    \REGION_{1b} & := \bigcap_{\omega \in [0, 1]} \left\{ \left. (R_1, R_2) \in \mathbb{R}^2_{\geq 0} \right| R_1 + \omega R_2 \leq \BOUND_{1b}(\omega)
                        \right\}  \label{eq:New-Bound1} \\
    \REGION_{1c} &:=  \bigcap_{\substack{\mu \in [0, \omega] \\ \omega \in [0, 1] }} \left\{\left. (R_1, R_2) \in
    \mathbb{R}^2_{\geq 0} \right| (1+\mu)R_1 + \omega R_2 \leq \BOUND_{1c}(\omega, \mu)
                        \right\} \label{eq:New-Bound2} 
  \end{align}
\end{subequations}
where
\begin{subequations}\label{eq:weighted}
  \begin{align}
     &\BOUND_{1a}(\omega) = \MEXPP\rN_{11} +
                        \omega \MEXP{\rN_{21} - \rN_{11}}^+ + \sum_{l=1}^q \left[\omega \beta_1(l) - \alpha_1(l)
                           \right]^+,
 \label{eq:weighted:a} \\
       &\BOUND_{1b}(\omega) \leq (1-\omega) \MEXPP\rN_{11} + \omega \MEXPP\rN_{21}  + \sum_{l=1}^q \left[ \omega \gamma_1(l) -
                               \alpha_1(l)\right]^+ + \omega\sum_{l=1}^q \max\big( \Prob{\rN_{11}-\rN_{21} \geq l},
                              \Prob{\rN_{12} \geq l} \big), \label{eq:weighted:b} \\
        & \BOUND_{1c}(\omega, \mu) = \MEXPP\rN_{11} + \omega\MEXP{ \rN_{21} - \rN_{11}}^+ + \sum_{l=1}^q \left[ \omega \gamma_1(l) -
      \alpha_1(l)\right]^+ +  \sum_{l=1}^q \max\left(  \mu \Prob{\rN_{11} \geq l},
      \omega\Prob{\rN_{12} \geq l} \right), \label{eq:weighted:c}
  \end{align}
\end{subequations}
and
\begin{subequations} \label{eq:abc}
  \begin{align}
    \alpha_1(l) &= \Prob{\rN_{21} \geq l } - \Prob{\rN_{21} - \rN_{11} \geq l }, \label{eq:alpha} \\
    \beta_1(l) & =  \big[ \Prob{\rN_{22} \geq l } - \Prob{\rN_{21} - \rN_{11} \geq l } \big]^+\label{eq:beta} \\
    \gamma_1(l) & = \big[ \Prob{\rN_{22} - \rN_{12}\geq l } - \Prob{\rN_{21}- \rN_{11} \geq l } \big]^+. \label{eq:gamma}
  \end{align}
\end{subequations}
By swapping subscripts $1$ and $2$ in~\eqref{eq:abc}, we can define
$\alpha_2(l)$, $\beta_2(l)$, and $\gamma_2(l)$ accordingly. For example, $\alpha_2(l) := \Prob{\rN_{12} \geq l } - \Prob{\rN_{12} - \rN_{22} \geq l }$. In turn, we can define $\BOUND_{2j}$ and $\REGION_{2j}$ for
$j \in \{a, b, c\}$ by exchanging subscripts $1$ with $2$ in~\eqref{eq:bounds} and~\eqref{eq:weighted}.

\begin{theorem}\label{thm:1}
  The capacity region $\mathcal{C}$ of the two-user interference channel~\eqref{eq:sys} is contained in
  $\REGION_{1a}\cap \REGION_{1b}\cap \REGION_{1c}\cap \REGION_{2a}\cap \REGION_{2b}\cap \REGION_{2c}$.
\end{theorem}

To help understand the outer bound, we make following remarks.
\begin{remark}
  Each region $\REGION_{kj}$, where $k\in\{1,2\}$ and $j\in\{a,b,c\}$, is a convex region characterized by a set of
  weighted sum-rate bounds, which is associated with each $\BOUND_{kj}$. Although the intersections are done among infinite numbers of $\omega$ and $\mu$ in~\eqref{eq:bounds}, it is not difficult to see that only finite numbers of them are
  necessary. Therefore, each $\REGION_{kj}$ is a polytope. In the remaining discussion,
  we refer to the weighted bound of form $R_1 + \omega R_2 \leq \BOUND_{1a}(\omega)$ as bound 
  $\BOUND_{1a}(\omega)$, and $(1+\mu) R_1 + \omega R_2 \leq \BOUND_{1c}(\omega, \mu)$ as bound $\BOUND_{1c}(\omega,
  \mu)$, \ETAL, for each fixed $\omega$ and $\mu$.
Furthermore, on the boundary of each region
  $\REGION_{1j}$ (inside the first orthant), $R_1$ has larger weight than $R_2$, which corresponds to a situation where rate of user~1 is preferred to that of user~2. By symmetry,
  similar interpretation can be made to each region $\REGION_{2j}$, where user~2 is the preferred one.
\end{remark}
\begin{remark}
  By setting each $\{\rN_{tr}[i]\}$, $t,r=1,2$, to a constant, say $n_{tr}$, the new outer bound recovers the capacity
  region of its deterministic counterpart, which consists of all positive rate pairs satisfying~\cite{Gamal82IFC, Bresler08IFC}:
  \begin{subequations} \label{eq:det}
    \begin{align}
      R_r &\leq n_{rr} \quad r = 1,2 \label{eq:det:a} \\
      R_1 + R_2 &\leq \max(n_{11}, n_{21}) + (n_{22} - n_{21})^+ \label{eq:det:b}  \\
      R_1 + R_2 &\leq \max(n_{22}, n_{12}) + (n_{11} - n_{12})^+  \label{eq:det:c} \\
      R_1 + R_2 &\leq \max((n_{11}-n_{21})^+, n_{12}) + \max((n_{22}-n_{12})^+, n_{21}) \label{eq:det:d} \\
      2R_1 + R_2 &\leq  \max(n_{11}, n_{12}) +  (n_{11}-n_{21})^+ + \max((n_{22}-n_{12})^+, n_{21}) \label{eq:det:e} \\
      R_1 + 2R_2 &\leq \max(n_{22}, n_{21}) + (n_{22}-n_{12})^+ +\max((n_{11}-n_{21})^+, n_{12}).\label{eq:det:f}
    \end{align}
  \end{subequations}
We leave the detailed proof in Appendix~\ref{app:det}. Table~\ref{tab:1} is a brief summary of the proof and it also
highlights the relation between each region $\REGION_{kj}$, $k=1,2$ and $j\in\{a,b,c\}$, and each constraint in~\eqref{eq:det}. 
We will look into more sophisticated examples in Section~\ref{sec:LEC}. 
\begin{table}[h]
  \centering
  \begin{tabular}{|c|c|}
    \hline Theorem~\ref{thm:1} & Constraint in~\eqref{eq:det} \\
    \hline $\BOUND_{1a}(\omega)$ and $\BOUND_{2a}(\omega)$ with $\omega = 0$ & \eqref{eq:det:a}\\
    \hline $\BOUND_{1a}(\omega)$ with $\omega = 1$ & \eqref{eq:det:b} \\
    \hline $\BOUND_{2a}(\omega)$ with $\omega = 1$ & \eqref{eq:det:c} \\
    \hline $\BOUND_{1b}(\omega)$ or $\BOUND_{2b}(\omega)$ with $\omega = 1$ & \eqref{eq:det:d} \\
    \hline $\BOUND_{1c}(\omega, \mu)$ with $\omega = 1, \mu = 1$ & \eqref{eq:det:e} \\
    \hline $\BOUND_{2c}(\omega, \mu)$ with $\omega = 1, \mu = 1$ & \eqref{eq:det:f} \\
    \hline
  \end{tabular}
  \caption{A summary of how to obtain capacity region of the deterministic IC via Theorem~\ref{thm:1}. It also shows
    how each bound $\BOUND_{kj}$, for $k=1,2$ and $j=\{a, b, c\}$, relates to each constraint of~\eqref{eq:det}.}
  \label{tab:1}
\end{table}
\end{remark}
\begin{remark}
  Regions $\REGION_{1a}$ and $\REGION_{2a}$ are actually capacity regions of Z-IC with $\rN_{21}\equiv
  0$ and $\rN_{12}\equiv 0$, respectively~\cite{ZhuGuo2011}. The remaining
  four are new.
\end{remark}

\section{More Discussion: Layered Erasure Cases} \label{sec:LEC}
In this section, we will investigate Theorem~\ref{thm:1}  under several special situations, which include cases where
Theorem~\ref{thm:1} is tight as well as some open cases. Inspired by classification done in~\cite{Vahid2015IC} for
$q=1$, we define following three cases:
\begin{enumerate}
\item Stochastically strong interference: $\forall l \in \{1,\dots,q\}$,
\begin{subequations}\label{eq:strong:a}
  \begin{align}
    \Prob{\rN_{12} \geq l} &\geq \Prob{\rN_{11} \geq l} \NONUM \\
    \Prob{\rN_{21} \geq l} &\geq \Prob{\rN_{22} \geq l}; \NONUM 
  \end{align}
\end{subequations}
\item Stochastically weak interference:  $\forall l \in \{1,\dots,q\}$,
  \begin{subequations}\label{eq:veryweak:a}
    \begin{align}
      \Prob{\rN_{11} - \rN_{21} \geq l } &\geq \Prob{\rN_{12} \geq l} \NONUM \\
      \Prob{\rN_{22} - \rN_{12} \geq l } &\geq \Prob{\rN_{21} \geq l};  \NONUM                                   
    \end{align}
  \end{subequations}
\item Stochastically moderate interference: $\forall l \in \{1,\dots,q\}$,
  \begin{subequations}\label{eq:moderate:a}
    \begin{align}
      \Prob{\rN_{11} \geq l } &> \Prob{\rN_{12} \geq l} > \Prob{\rN_{11} -
                                \rN_{21} \geq l } \label{eq:moderate:c1} \\
      \Prob{\rN_{22} \geq l} &> \Prob{\rN_{21} \geq l} > \Prob{\rN_{22} - \rN_{12} \geq l }. \label{eq:moderate:c2} 
    \end{align}
  \end{subequations}
\end{enumerate}

The first case can be interpreted as following: from the viewpoint of any given layer $l$, signal reaches to the undesired user more often
than the desired one, which is similar to the strong interference channels in usual sense~\cite{Costa87IFC}. Both of the
other two cases implies that $\Prob{\rN_{12} \geq l}  \leq \Prob{\rN_{11} \geq l}$ and $\Prob{\rN_{21} \geq l}  \leq
\Prob{\rN_{22} \geq l}$ for any $l$. Therefore, they can be
interpreted as cases where signal reaches to the desired user more often than the undesired one from the viewpoint of each
layer. Hence, they both fall into weak interference category in usual sense. 

In this section, we will show that the new outer bound actually coincides with the capacity regions for the first two
cases. However, for the moderate interference case, whether Theorem~\ref{thm:1} is tight still remains open. Note that
general layered erasure channel can be much more complicated so that it can be none of these three cases. We will
conclude this section with some discussion about general cases.

\subsection{Stochastically Strong Interference} \label{sec:strong}
For strong interference, it is well known that the capacity region is the same as capacity region of compound multi-access channel
at receivers~1 and 2~\cite{Costa87IFC}, \IE, 
\begin{align}\label{eq:strong:b}
  \left\{ (R_1, R_2) \in \mathbb{R}^2_{\geq 0} \left|
  \begin{array}{rcl}
    R_i &\leq& \MEXPP\rN_{rr} \quad r = 1,2\\
    R_1 + R_2 &\leq& \max\big( \MEXPP\max(\rN_{11}, \rN_{21}), \MEXPP\max(\rN_{22}, \rN_{12}) \big) 
  \end{array}
\right\}. \right.
\end{align}
From Theorem~\ref{thm:1}, we see that all summation terms vanish in $\BOUND_{kj}$, $k=1,2$ and $j\in \{a,b,c\}$ under
stochastically strong interference assumption. Therefore, it is not difficult to see that the outer bound becomes
$\REGION_{1a}\cap\REGION_{2a}$, which coincides with the region defined by~\eqref{eq:strong:b}.

\subsection{Stochastically Weak Interference}\label{sec:veryweak}
\begin{figure}[h]
  \centering
  \includegraphics[width=.5\textwidth]{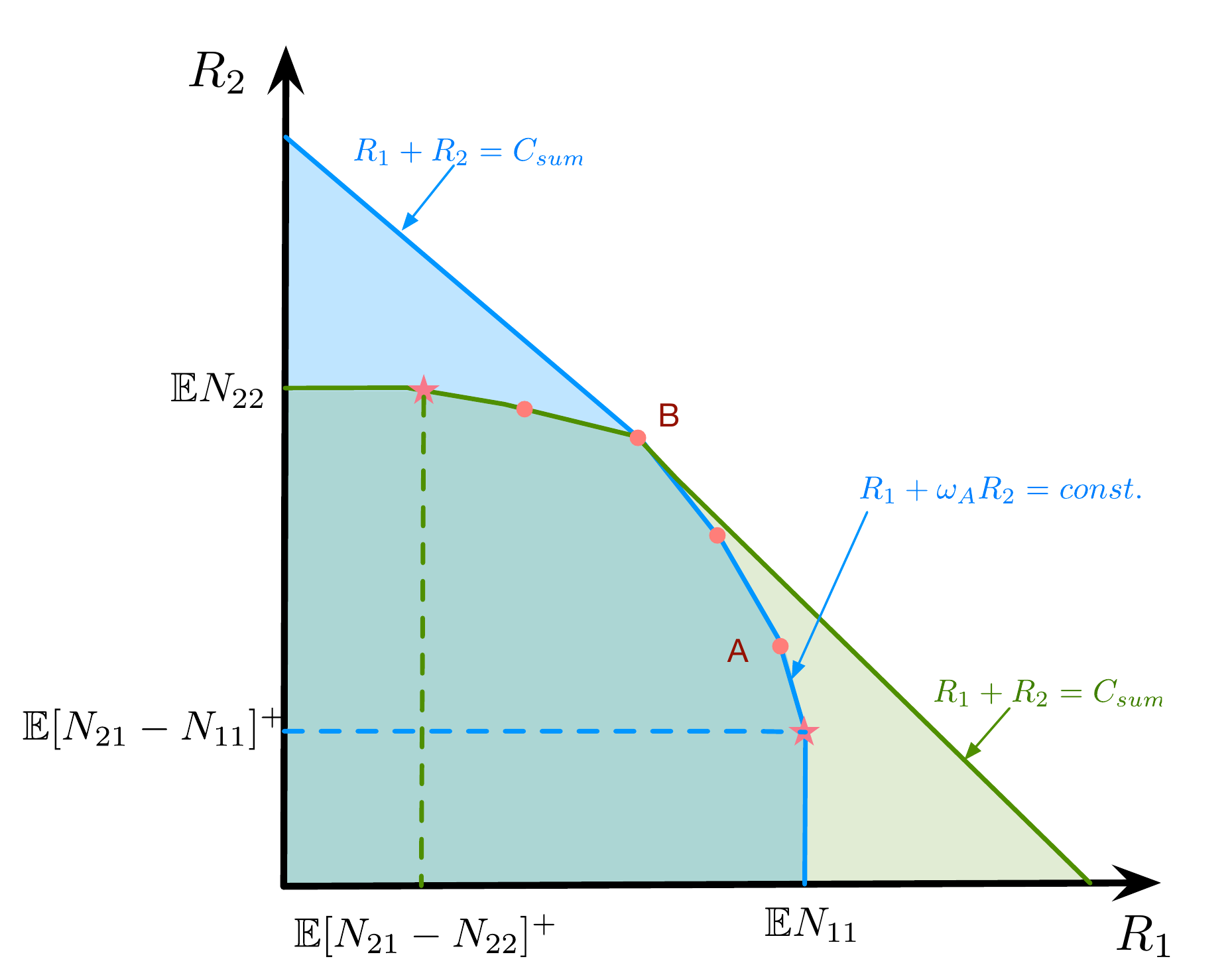}
  \caption{Sketch of capcity region of a weak interference channel. Regions with blue and green shadow are
    $\REGION_{1b}$ and $\REGION_{2b}$, respectively. Each of them is a polytope and their intersection is the capacity
    region. Point B is the intersection between boundaries of the two regions and it achieves the sum-capacity.}
  \label{fig:cap}
\end{figure}

\begin{theorem}\label{thm:2}
  For interference channel which satisfies condition~\eqref{eq:veryweak:a} $\forall$ $l \in \{1,
  \dots, q\}$, the capacity region is characterized by
  Theorem~\ref{thm:1}. In particular, it coincides with $\REGION_{1b} \cap \REGION_{2b}$ and sum-capacity is given by
  \begin{align}
    C_{sum} = \MEXP{\rN_{22} - \rN_{12}}^+ + \MEXP{\rN_{11}-\rN_{21}}^+. \label{eq:veryweak:sumcap}
  \end{align}
\end{theorem}
\begin{IEEEproof}
  We start with the converse, which makes construction of achievable schemes more intuitive. Since \eqref{eq:veryweak:a} holds for
  any $l$, we can
  simplify $\BOUND_{1b}(\omega)$ as
  \begin{align}
    \BOUND_{1b}(\omega) &= (1-\omega) \MEXPP\rN_{11} + \omega \MEXPP\rN_{21}  + \sum_{l=1}^q \left[ \omega \gamma_1(l) -
                               \alpha_1(l)\right]^+ + \omega\sum_{l=1}^q \Prob{\rN_{11}-\rN_{21} \geq l} \NONUM \\
  & = (1-\omega) \MEXPP\rN_{11} + \omega \MEXPP\rN_{21}  + \sum_{l=1}^q \left[ \omega \gamma_1(l) -
                               \alpha_1(l)\right]^+ + \omega \MEXP{\rN_{11}-\rN_{21}}^+ \NONUM \\
 & = \MEXPP\rN_{11} + \omega \MEXP{\rN_{21}-\rN_{11}}^+ + \sum_{l=1}^q \left[ \omega \gamma_1(l) -
                               \alpha_1(l)\right]^+. \label{eq:veryweak:R1b}
\end{align}
With \eqref{eq:veryweak:a} and \eqref{eq:beta}, we have $\beta_1(l) \geq \gamma_1(l)$. Therefore $\BOUND_{1b}(\omega) \leq
\BOUND_{1a}(\omega)$ $\forall \omega \in [0,1]$, which implies $\REGION_{1b} \subset \REGION_{1a}$.
For $\BOUND_{1c}(\omega, \mu)$, we have
\begin{align}
    \BOUND_{1c}(\omega, \mu) &= \MEXPP\rN_{11} + \omega\MEXP{ \rN_{21} - \rN_{11}}^+ + \sum_{l=1}^q \left[ \omega \gamma_1(l) -
      \alpha_1(l)\right]^+ \nonumber \NONUM \\
      &\quad +  \sum_{l=1}^q \big[ \max\big(  \mu \Prob{\rN_{11} \geq l},
      \omega\Prob{\rN_{12} \geq l} \big) - \mu \Prob{\rN_{11} \geq l} + \mu \Prob{\rN_{11} \geq l} \big] \NONUM \\
    & = (1+\mu)\MEXPP\rN_{11} + \omega\MEXP{ \rN_{21} - \rN_{11}}^+ + \sum_{l=1}^q \left[ \omega \gamma_1(l) -
      \alpha_1(l)\right]^+ \nonumber \\
    & \quad + \sum_{l=1}^q \left[ \omega\Prob{\rN_{12} \geq l} - \mu \Prob{\rN_{11} \geq l}\right]^+. 
  \end{align}
Therefore, bound $\BOUND_{1c}(\omega, \mu)$ can be rewritten as
\begin{align}
  R_1 + \frac{\omega}{1+\mu}R_2 &\leq \MEXPP\rN_{11} + \frac{\omega}{1+\mu}\MEXP{ \rN_{21} - \rN_{11}}^+ + \sum_{l=1}^q \left[ \frac{\omega}{1+\mu} \gamma_1(l) -
      \frac{1}{1+\mu}\alpha_1(l)\right]^+ \nonumber \\
    & \quad + \sum_{l=1}^q \left[ \frac{\omega}{1+\mu}\Prob{\rN_{12} \geq l} - \frac{\mu}{1+\mu} \Prob{\rN_{11} \geq l}\right]^+. \label{eq:veryweak:R1c}
\end{align}
Comparing with~\eqref{eq:veryweak:R1b}, we see that the right-hand side~(RHS) of~\eqref{eq:veryweak:R1c} is greater or equals to
$\BOUND_{1b}(\frac{\omega}{1+\mu})$ for each fixed $\omega$ and $\mu$. Therefore, $\REGION_{1b} \subset
\REGION_{1c}$. Hence, we have $\REGION_{1b} \subset
\REGION_{1a} \bigcap \REGION_{1c}$.  By symmetry, we can argue that
$\REGION_{2b} \subset \REGION_{2a} \cap \REGION_{2c}$. Therefore, the outer bound in Theorem~\ref{thm:1} equals
to $\REGION_{1b} \cap \REGION_{2b}$ under the stochastically weak interference assumption.

Let $\omega = 1$ in~\eqref{eq:veryweak:R1b}, we obtain
\begin{align}
  R_1 + R_2 &\leq \MEXPP\max\big(\rN_{11}, \rN_{21}\big)  + \sum_{l=1}^q \left[ \gamma_1(l) -
                               \alpha_1(l)\right]^+. \label{eq:veryweak:sumR}
\end{align}
With \eqref{eq:gamma} and~\eqref{eq:alpha}, we have
\begin{align}
  \gamma_1(l) - \alpha_1(l)  &= \big[ \Prob{\rN_{22} - \rN_{12} \geq l } - \Prob{\rN_{21} -
                                                \rN_{11} \geq l} \big]^+ - \Prob{\rN_{21} \geq l} + \Prob{\rN_{21} - \rN_{11} \geq l } \NONUM  \\
           &=  \max\big( \Prob{\rN_{22} - \rN_{12} \geq l }, \Prob{\rN_{21} -
                                                \rN_{11} \geq l} \big) - \Prob{\rN_{21} \geq l}. \NONUM
\end{align}
With condition~\eqref{eq:veryweak:a}, we have
$\Prob{\rN_{22} - \rN_{12} \geq l } \geq \Prob{\rN_{21} \geq l} \geq \Prob{\rN_{21} - \rN_{11} \geq l}$. Therefore,
\begin{align}
          \gamma_1(l) - \alpha_1(l)  & = \Prob{\rN_{22} - \rN_{12} \geq l } - \Prob{\rN_{21} \geq l}. \label{eq:gamma_vs_alpha}
\end{align}
From condition~\eqref{eq:veryweak:a} and equation~\eqref{eq:gamma_vs_alpha}, we observe that
that $\gamma_1(l) - \alpha_1(l) \geq 0$ for any $l$. Substitute~\eqref{eq:gamma_vs_alpha}
into~\eqref{eq:veryweak:sumR}, we have $R_1 + R_2 \leq C_{sum}$.

By symmetry, we have shown that $\BOUND_{1b}(1) = \BOUND_{2b}(1) = C_{sum}$. In fact, it is not difficult to see that
boundaries of $\REGION_{1b}$ and $\REGION_{2b}$ intersects inside the first orthant at point
$(R_1, R_2)=$$(\MEXP{\rN_{11}-\rN_{21}}^+$
$,$
$\MEXP{\rN_{22}
  -
  \rN_{12}}^+)$. Indeed, this point can be achieved by treating interference as noise at both receivers. We illustrate
$\REGION_{1b}$
and $\REGION_{2b}$
in Fig.~\ref{fig:cap} by shadowing them with blue and green, respectively. To show the overlapped region is achievale,
it is sufficient to construct a coding scheme for each extreme point. Let us focus on one of them, say, point A in
Fig.~\ref{fig:cap}.

\eqref{eq:veryweak:R1b} holds some insights about the coding schemes.
Note that $\big(\MEXPP\rN_{11}, \MEXP{\rN_{21}-\rN_{11}}^+\big)$ is always on the boundary of $\REGION_{1b}$, which
marks by $\bigstar$ in Fig~\ref{fig:cap}. If user~1 and user~2 transmit their information at rate $\MEXPP\rN_{11}$ and
$\MEXP{\rN_{21}-\rN_{11}}^+$, respectively, user~1's rate is achievable by decoding and canceling user~2's message
completely. One the other hand, user~2 should not have issue to decode its message at rate of
$\MEXP{\rN_{21}-\rN_{11}}^+$ by treating user~1's signal as noise, because $\MEXP{\rN_{21}-\rN_{11}}^+ \leq
\MEXP{\rN_{22} - \rN_{12}}^+$ by weak interference assumption. If user~2 would like to send at higher rate, it faces a
tradeoff by generating more interference, which, consequently, will reduce user~1's rate. This type of tradeoff is captured by the summation term
in~\eqref{eq:veryweak:R1b}. Therefore an achievable scheme for point A can be constructed as following. Both users
generate their codebooks according to the distribution of q-size random vector whose elements are all
i.i.d. $\Ber{1/2}$ random variables. Here, $\Ber{p}$ denotes Bernoulli distribution with probability $p$ taking value $1$
and probability $1-p$ taking value $0$. To finalize the coding scheme, we only need to determine the allocation of
layers for private and common
messages, respectively, in the spirit of Han-kobayashi~(HK) scheme. 

Suppose that the right boundary segment of point A is on a line of form $R_1 + \omega_{A}R_2=const.$ Let us define
two subsets of $\{1, \dots, q\}$: $\mathcal{L}_p := \{ l | \omega_A\gamma_1(l) \geq \alpha_1(l) \}$ and 
$\mathcal{L}_c := \{ l | \omega_A\gamma_1(l) < \alpha_1(l) \}$. Thus, we can write the coordinate of point
A as
\begin{align}
  \left( \MEXPP\rN_{11} - \sum_{l \in \mathcal{L}_p} \alpha_1(l) \, , \, \MEXP{\rN_{21}-\rN_{11}}^+ + \sum_{l \in
  \mathcal{L}_p} \gamma_1(l)\right).
\end{align}
In terms of coding scheme, we allocate all layers of user~1 for private message. For user~2, we allocate layers
in $\mathcal{L}_p$
to carry private message and layers in $\mathcal{L}_c$ to carry common message. 
At receiving sides, receiver~1 decodes and
removes common message of user~2 before decoding its own message. By evaluating the corresponding mutual information, one
can verify that user~1 can achieve transmit rate at $\MEXPP\rN_{11} - \sum_{l
  \in \mathcal{L}_p} \alpha_1(l)$. On the other hand, we have
\begin{align}
  & \MEXP{\rN_{21}-\rN_{11}}^+ + \sum_{l \in
  \mathcal{L}_p} \gamma_1(l)  \nonumber \\
  = &\MEXP{\rN_{21}-\rN_{11}}^+ + \sum_{l=1}^q \gamma_1(l) - \sum_{l \in
  \mathcal{L}_c} \gamma_1(l) \NONUM\\
 = &\sum_{l=1}^q \max\big( \Prob{\rN_{22} - \rN_{12} \geq l }, \Prob{\rN_{21} -
                                                \rN_{11} \geq
                   l} \big)  - \sum_{l \in
  \mathcal{L}_c} \gamma_1(l) \NONUM \\
  = &\sum_{l=1}^q \Prob{\rN_{22} - \rN_{12} \geq l } - \sum_{l \in
  \mathcal{L}_c} \gamma_1(l) \label{eq:veryweakachieve:t3} \\
 & \geq  \MEXP{\rN_{22} - \rN_{12}}^+ \label{eq:veryweakachieve:t4}
\end{align}
where \eqref{eq:veryweakachieve:t3} is due
to weak interference assumption~\eqref{eq:veryweak:a}. By \eqref{eq:veryweakachieve:t4}, we see that user~2 can decode
its own message at rate $\MEXP{\rN_{21}-\rN_{11}}^+ + \sum_{l \in
  \mathcal{L}_p} \gamma_1(l)$ by treating interference from user~1 as noise.
\end{IEEEproof}

\subsection{Stochastically Moderate Interference}
Whether the outer bound given by Theorem~\ref{thm:1} is also tight or not for moderate interference is not clear. But
with condition~\eqref{eq:moderate:a}, we can simply the bound further:
\begin{proposition}\label{pro:1}
  For moderate interference channel, which satisfies condition~\eqref{eq:moderate:a} $\forall l$, we have
  \begin{subequations}\label{eq:moderate:weight}
    \begin{align}
      &\BOUND_{1a}(\omega) = \MEXPP\rN_{11} +
        \omega \MEXP{\rN_{21} - \rN_{11}}^+ + \sum_{l=1}^q \left[\omega \beta_1(l) - \alpha_1(l)
        \right]
        \label{eq:moderate:w1}\\
      &\BOUND_{1b}(\omega) \leq (1-\omega) \MEXPP\rN_{11} + \omega \big( \MEXPP\rN_{21} + \MEXPP\rN_{12} \big)\label{eq:moderate:w2} \\
      & \BOUND_{1c}(\omega, \mu) = \MEXPP\rN_{11} + \omega\MEXP{ \rN_{21} - \rN_{11}}^+ +  \sum_{l=1}^q \max\left(  \mu \Prob{\rN_{11} \geq l},
        \omega\Prob{\rN_{12} \geq l} \right), \label{eq:moderate:w3}
    \end{align}
   and $\BOUND_{2a}$, $\BOUND_{2b}$ and $\BOUND_{2c}$ are in  similar forms with subscripts $1$ and $2$ exchanged, accordingly.
  \end{subequations}
\end{proposition}

The proof is similar as reverse part of Theorem~\ref{thm:2}, except that with
condition~\eqref{eq:moderate:a} we have $\gamma_k(l) < \alpha_k(l)$, for $k=1,2$ and $\forall l \in \{1,\dots,q\}$.

With $q=1$, Proposition~\ref{pro:1} does not improve results for sum-rate obtained in~\cite{Vahid2015IC}. However, it does
provide a slightly better outer bound for the whole capacity region. To see this, we simplify the notation for
symmetric single-layered case, \IE~$q=1$ as following. $\Prob{\rN_{11} = 1} = \Prob{\rN_{22} = 1} = p_d$ and
$\Prob{\rN_{21}=1} = \Prob{\rN_{12} = 1} = p_c$. The moderate interference assumption~\eqref{eq:moderate:a} becomes
$p_d > p_c > \frac{p_d}{1+p_d}$, or, equivalently, $p_d p_c > p_d - p_c > 0$. With these notations, we have
\begin{align}
\alpha_1(1) & = \alpha_2(1)  = p_dp_c \NONUM \\
\beta_1(1) & = \beta_2(1)  = p_d -p_c +  p_dp_c \NONUM
\end{align}

For $k=1,2$, consider $\BOUND_{ka}(0)$, $\BOUND_{ka}\left(\frac{\alpha_1(1)}{\beta_1(1)}\right)$,
$\BOUND_{kb}\left(1\right)$, and $\BOUND_{kc}\left(1,
  \frac{p_c}{p_d}\right)$ accordingly. We obtain an outer bound as
\begin{align}\label{eq:moderate:q1}
  \left\{ (R_1, R_2) \in \mathbb{R}^2_{\geq 0} \left| 
  \begin{array}[l]{rcl}
    R_i &\substack{(a)\\\leq}& p_d  \quad i=1,2 \\
    R_1 + \frac{p_d p_c}{p_d -p_c + p_dp_c} R_2 &\substack{(b)\\\leq} &p_d + \frac{p_d p_c}{p_d - p_c+p_dp_c} p_c(1-p_d) \\
    \frac{p_d p_c}{p_d -p_c - p_dp_c}  R_1 + R_2 &\substack{(c)\\\leq} &p_d + \frac{p_d p_c}{p_d - p_c+p_dp_c} p_c(1-p_d) \\
    R_1 + R_2 &\substack{(d)\\\leq} & 2p_c \\
    R_1 + \frac{p_d}{p_d + p_c} R_2 &\substack{(e)\\\leq} &p_d + \frac{p_d}{p_d + p_c} p_c(1-p_d) \\
    \frac{p_d}{p_d + p_c}  R_1 + R_2 &\substack{(f)\\\leq} &p_d + \frac{p_d}{p_d + p_c} p_c(1-p_d) 
  \end{array}
\right\} \right. 
\end{align}
Note that conditions $(a)$--$(d)$ are shown in~\cite{Vahid2015IC} and none of them is redundant when $p_d p_c > p_d - p_c >
0$. We will show that in some condition $(e)$ is stronger than $(b)$. In fact, we observe that lines $ R_1 + \frac{p_d p_c}{p_d - p_c+p_dp_c} R_2 = p_d + \frac{p_d p_c}{p_d -
  p_c + p_dp_c} p_c(1-p_d)$ and $ R_1 + \frac{p_d}{p_d + p_c} R_2 = p_d + \frac{p_d}{p_d + p_c} p_c(1-p_d)$ intersect at
$(p_d, p_c(1-p_d))$. Regarding the slopes of these two lines, $\frac{p_d}{p_d + p_c} > \frac{p_d p_c}{p_d - (1-p_d)p_c}$
under the condition of $p_d-p_c
> p_c^2$. In other word, when $p_d-p_c > p_c^2$, $(b)$ becomes redundant to $(e)$. Note that $p_d-p_c > p_c^2$
can happen under moderate interference assumption. 

\subsection{General Cases}
One can evaluate Theorem~\ref{thm:1} under all kinds of different conditions beyond what we have discussed above. For example, one
can mixed conditions of weak, moderate, strong interference between the two users. Moreover, in above three situations,
we require \eqref{eq:strong:a}, \eqref{eq:veryweak:a}, or \eqref{eq:moderate:a} hold for all $l$. One can also come up a
case where conditions of \eqref{eq:strong:a}, \eqref{eq:veryweak:a}, and \eqref{eq:moderate:a} are mixed across
different layers. Therefore, structures of general layered erasure channels could be very complicated. However, we
believe the limitation of Theorem~\ref{thm:1} roots on our limited understanding of the moderate interference with
$q=1$. In particular, we conjecture that
\begin{conjecture}
  Theorem~\ref{thm:1} characterizes the capacity region of general layered erasure channels, which satisfy
  condition that there does not exist $l \in \{1, \dots, q\}$ such that  either  \eqref{eq:moderate:c1} or
  \eqref{eq:moderate:c2} holds.
\end{conjecture}

\section{Proof of Theorem~\ref{thm:1}}\label{sec:proof}

\subsection{Preliminary Lemmas}
The proof relies on following two lemmas. 
\begin{lemma}
\label{lm:marton}
Consider $n$ uses of a memoryless channel described by random transformation $\Probb_{\rY,\rZ,\rT|\rX,\rS}$.  Let $\rvX^n$ and $\rvS^n$ denote the independent input and state sequences, respectively. 
Then for any $\mu_2 \geq \mu_1 \geq 0$, 
  \begin{align}
    \mu_1\MI{\rvX^n; \rvZ^n | \rvT^n, \rvS^n} &- \mu_2\MI{\rvX^n;
      \rvY^n | \rvT^n, \rvS^n} \nonumber \\
&\leq \sum_{i=1}^n \Big[ \mu_1\MI{\rX_i; \rZ_i | \rvZ^{i-1},
    \rvY_{i+1}^n,  \rvT^n,
   \rvS^n} 
- \mu_2\MI{\rX_i; \rY_{i} | \rvZ^{i-1},
    \rvY_{i+1}^n,  \rvT^n, \rvS^n}\Big]. \label{eq:newMarton}
  \end{align}
\end{lemma}

It is a Marton-style expansion of mutual information, which essentially converts
a multi-letter mutual information difference into a single-letter one. For the proof, please refer to~\cite[Appendix A]{ZhuGuo2011}~\footnote{In
  \cite{ZhuGuo2011}, the lemma is in a form of $\mu_2 = 1$ and $\mu_1 \in [0,1]$. Lemma~\ref{lm:marton} here is a trivial
  extension.}.

\begin{lemma} \label{lm:mac}
  Suppose that $\SEQN{\rvW}$ and $\SEQN{\rvX}$ are two independent arbitrary random processes taking value in $\mathbb{F}^q_2$. Let
    $\SEQN{\rN_1}$ and $\SEQN{\rN_2}$ be two i.i.d fading processes taking value in $\{0, \dots, q\}$. Then we have
    \begin{align}
      \CMI{\SEQN{\rvW}, \SEQN{\rvX}; \SEQN{\rvW^{\rN_1} \oplus \rvX^{\rN_2}}}{\SEQN{\rN_1}, \SEQN{\rN_2}}  \leq n\MEXPP\rN_1 + \CMI{\SEQN{\rvX}; \SEQN{\rvX^{(\rN_2-\rN_1)^+}}}{\SEQN{\rN_1}, \SEQN{\rN_2}} \label{eq:lm:erasure}
    \end{align}
\end{lemma}
\begin{IEEEproof}
  \begin{align}
   & \CMI{\SEQN{\rvW}, \SEQN{\rvX}; \SEQN{\rvW^{\rN_1} \oplus \rvX^{\rN_2}}}{\SEQN{\rN_1}, \SEQN{\rN_2}}  \NONUM \\
= &\CENT{\SEQN{\rvW^{\rN_1}\oplus\rvX^{\rN_2}}}{\SEQN{\rN_1},\SEQN{\rN_2}} \NONUM \\
\leq & \CENT{\SEQN{\rvW^{\rN_1}\oplus \rvWt^{\rN_1} \oplus\rvX^{\rN_2}}}{\SEQN{\rN_1},\SEQN{\rN_2}}  \NONUM
  \end{align}
where $\SEQN{\rvWt}$ be an i.i.d. random sequence independent of all other random processes and each element of
  $\rvWt[i]$ is an independent $\Ber{1/2}$ random variable. Note that $\SEQN{\rvWt \oplus\rvW}$ has same distribution as
  $\rvWt$ and it is also independent of $\SEQN{\rvX}$. Let $\rvWb[i] = \rvWt[i] \oplus\rvW[i]$. Then
  \begin{align}
   & \CMI{\SEQN{\rvW}, \SEQN{\rvX}; \SEQN{\rvW^{\rN_1} \oplus \rvX^{\rN_2}}}{\SEQN{\rN_1}, \SEQN{\rN_2}}  \NONUM \\
\leq & \CENT{\SEQN{\rvWb^{\rN_1} \oplus\rvX^{\rN_2}}}{\SEQN{\rN_1},\SEQN{\rN_2}} \NONUM \\
  =  & \CMI{\SEQN{\rvWb}, \SEQN{\rvX}; \SEQN{\rvWb^{\rN_1} \oplus \rvX^{\rN_2}}}{\SEQN{\rN_1}, \SEQN{\rN_2}} \NONUM \\
  = & \MEXPP\rN_1 + \CMI{\SEQN{\rvX}; \SEQN{\rvWb^{\rN_1} \oplus \rvX^{\rN_2}}}{\SEQN{\rN_1}, \SEQN{\rN_2}}
      \label{eq:lm:mac:proof1} \\
 =    & \MEXPP\rN_1 + \CMI{\SEQN{\rvX}; \SEQN{\rvX^{(\rN_2-\rN_1)^+}}, \SEQN{\rvWb^{\rN_1} \oplus \rvX^{\rN_1}}}{\SEQN{\rN_1},
     \SEQN{\rN_2}}. \label{eq:lm:mac:proof2}
  \end{align}
where~\eqref{eq:lm:mac:proof1} is due to chain rule and~\eqref{eq:lm:mac:proof2} leads us to \eqref{eq:lm:erasure} by noting that $\SEQN{\rvWb^{\rN_1} \oplus \rvX^{\rN_1}}$ and
  $\SEQN{\rvX^{(\rN_2-\rN_1)^+}}$ are independent of each other given $\left(\SEQN{\rN_1}, \SEQN{\rN_2}\right)$.
\end{IEEEproof}

Basically, Lemma~\ref{lm:mac} claims that for multi-access layered erasure channel, we can fix one input as $\SEQN{\rvWt}$
without reducing the sum-rate.

\subsection{Proof}
First, by letting $\rN_{12}[i]\equiv 0$, we obtained a z-interference channel, whose capacity region can serve as a
natural 
outer bound. Therefore,  bounds $\BOUND_{1a}(\omega)$ and $\BOUND_{2a}(\omega)$ are direct
consequence of \cite[Theorem 2]{ZhuGuo2011}. Thus, it is sufficient to show the
remaining four bounds. By symmetry, we only need to show bounds $\BOUND_{1b}$ and $\BOUND_{1c}$, respectively.

For shorthand notation, we define
$\rvN[i] = \big( \rN_{11}[i], \rN_{12}[i], \rN_{22}[i], \rN_{21}[i] \big)$.  $\forall \omega \in [0,1]$, applying Fano's inequality at receiver 1, we have
\begin{align}
  nR_1 - n\delta_n&\leq \CMI{\SEQN{\rvW}; \SEQN{\rvY}}{\SEQN{\rvN}}  \NONUM \\
          & = \CMI{\SEQN{\rvW}, \SEQN{\rvX}; \SEQN{\rvY}}{\SEQN{\rvN}} - \CMI{\SEQN{\rvX};
            \SEQN{\rvX^{\rN_{21}}}}{\SEQN{\rvN}} \label{eq:R1preCom:t1} \\
          & =  \omega \CMI{\SEQN{\rvW}, \SEQN{\rvX}; \SEQN{\rvY}}{\SEQN{\rvN}} + (1-\omega)\CMI{\SEQN{\rvW},
            \SEQN{\rvX}; \SEQN{\rvY}}{\SEQN{\rvN}}  \nonumber \\
       & \qquad - \CMI{\SEQN{\rvX}; \SEQN{\rvX^{\rN_{21}}}}{\SEQN{\rvN}} \NONUM \\
       & \leq \omega \CMI{\SEQN{\rvW}, \SEQN{\rvX}; \SEQN{\rvY}}{\SEQN{\rvN}} + n(1-\omega)\MEXPP\rN_{11} + (1-\omega)\CMI{
            \SEQN{\rvX}; \SEQN{\rvX^{(\rN_{21}-\rN_{11})^+}}}{\SEQN{\rvN}} \nonumber \\
        & \qquad - \CMI{\SEQN{\rvX}; \SEQN{\rvX^{\rN_{21}}}}{\SEQN{\rvN}} \label{eq:R1preCom} 
\end{align}
where \eqref{eq:R1preCom:t1}
is due to chain rule; in \eqref{eq:R1preCom}, we apply Lemma~\ref{lm:mac} by letting $\rN_1=\rN_{11}$ and $\rN_2 =
\rN_{21}$; and $\delta_n$ vanishes as $n \to \infty$. 

Apply Fano's inequality at receiver 2, we have
\begin{align}
  nR_2 - n\delta_n & \leq \CMI{\SEQN{\rvX}; \SEQN{\rvZ}}{\SEQN{\rvN}} \NONUM \\
  & =  \CMI{\SEQN{\rvW}, \SEQN{\rvX}; \SEQN{\rvZ}}{\SEQN{\rvN}} - \CMI{\SEQN{\rvW}; \SEQN{\rvW^{\rN_{12}}}}{\SEQN{\rvN}}
     \label{eq:R2preCom:t1}\\
 & = \MEXPP\rN_{12} + \CMI{\SEQN{\rvX}; \SEQN{\rvX^{(\rN_{22}-\rN_{12})^+}}}{\SEQN{\rvN}} -
    \CMI{\SEQN{\rvW}; \SEQN{\rvW^{\rN_{12}}}}{\SEQN{\rvN}} \label{eq:R2preCom} 
\end{align}
where~\eqref{eq:R2preCom:t1} is due to chain rule and in \eqref{eq:R2preCom}, we apply Lemma with $\rN_1=\rN_{12}$ and
$\rN_{2}=\rN_{22}$. 

By combining~\eqref{eq:R1preCom} and~\eqref{eq:R2preCom}, we can get a weighted bound:  
\begin{align}
  n(R_1 &+ \omega R_2 - (1+\omega)\delta_n) \nonumber \\
    & \leq n(1-\omega) \MEXPP\rN_{11} + n\omega \MEXPP\rN_{12} + A + B \label{eq:weightCom1}
\end{align}
where
\begin{align}
  A & = \omega\CMI{\SEQN{\rvX};
       \SEQN{\rvX^{(\rN_{22}-\rN_{12})^+}}}{\SEQN{\rvN}} - \CMI{\SEQN{\rvX};
       \SEQN{\rvX^{\rN_{21}}}}{\SEQN{\rvN}} \nonumber \\
    & \qquad +(1-\omega)\CMI{
            \SEQN{\rvX}; \SEQN{\rvX^{(\rN_{21}-\rN_{11})^+}}}{\SEQN{\rvN}} \NONUM \\
  B &= \omega \CMI{\SEQN{\rvW}, \SEQN{\rvX}; \SEQN{\rvY}}{\SEQN{\rvN}} -
    \omega \CMI{\SEQN{\rvW}; \SEQN{\rvW^{\rN_{12}}}}{\SEQN{\rvN}}. \label{eq:partB}
\end{align}
We will deal with $A$ and $B$ separated. 

Starting with $A$, let $\rM[i] = (\rN_{22}[i] - \rN_{12}[i])^+$ and $\rL[i] = (\rN_{21}[i]-\rN_{11}[i])^+$ for each $i$. $A$
can be rewritten as
\begin{align}
  A & = \omega\CMI{\SEQN{\rvX};
       \SEQN{\rvX^{\rM}}}{\SEQN{\rM}} - \CMI{\SEQN{\rvX};
       \SEQN{\rvX^{\rN_{21}}}}{\SEQN{\rN_{21}}}  +(1-\omega)\CMI{
            \SEQN{\rvX}; \SEQN{\rvX^{\rL}}}{\SEQN{\rL}}. \NONUM
\end{align}
We observe that given $\SEQN{\rvX}$, above mutual information terms \emph{only} depend on distributions of $\rM$, $\rN_{21}$, and $\rL$,
respectively. Therefore, we can replace those three i.i.d random processes with $\SEQN{\rMt}$,
$\SEQN{\rNt_{21}}$ and $\SEQN{\rLt}$, respectively, without changing the value of $A$, as long as they satisfy
\begin{align}
  \SEQN{\rMt} \sim \SEQN{\rM} \text{,   } \SEQN{\rNt_{21}} \sim \SEQN{\rN_{21}}, \text{  and  } \SEQN{\rLt} \sim
  \SEQN{\rL} \label{eq:coupling:condition}
\end{align}
where $\sim$ indicates the two objects on 
its two sides follow the same statistical law. In particular, we consider a construction of $\SEQN{\rMt}$, $\SEQN{\rNt}$
and $\SEQN{\rLt}$ as following.  Let $\{\rlambda[i]\}$ be an i.i.d. random process
with uniform distribution over interval $[0,1]$. We also assume $\{\rlambda[i]\}$ is also independent of other random
variables or vectors. For each $i$, define
\begin{subequations} \label{eq:alignment}
  \begin{align}
    \rMt[i] &= F_{\rM}^{-1}(\rlambda[i]) = F_{(\rN_{22}-\rN_{12})^+}^{-1}(\rlambda[i]) \NONUM \\
    \rNt_{21}[i] &= F_{\rN_{21}}^{-1}(\rlambda[i]) \NONUM \\
    \rLt[i] &= F_{\rL}^{-1}(\rlambda[i]) =  F_{ (\rN_{21}-\rN_{11})^+ }^{-1}(\rlambda[i])\NONUM
  \end{align}
\end{subequations}
where, for any random
variable $\rN$, let $F_{\rN}(n) = \Prob{\rN \leq n}$ denote its the cumulative distribution function and  its
pseudo-inverse is defined as $F_{\rN}^{-1}(v) = \inf\{u | F_{\rN}(u) \geq v \}$. One can verify that
condition~\eqref{eq:coupling:condition} is satisfied. Therefore, we can do the replacement safely. With~\eqref{eq:alignment}, we also see that $\rLt[i]
\leq \rNt_{21}[i]$ for any $i$. Hence, $\SEQN{\rvX}$---$\SEQN{\rvX^{\rNt_{21}}}$---$\SEQN{\rvX^{\rLt}}$ is a Markov chain.
Thus, $A$ can be further rewritten as
\begin{align}
  A & = \omega\CMI{\SEQN{\rvX};
       \SEQN{\rvX^{\rMt}}}{\SEQN{\rlambda}} - \CMI{\SEQN{\rvX};
       \SEQN{\rvX^{\rNt_{21}}} }{ \SEQN{\rlambda} }  +(1-\omega)\CMI{
            \SEQN{\rvX}; \SEQN{\rvX^{\rLt}}}{\SEQN{\rlambda}}. \NONUM  \\
     & = \omega\CMI{\SEQN{\rvX};
       \SEQN{\rvX^{\rMt}}}{\SEQN{\rlambda}}  -\omega \CMI{
            \SEQN{\rvX}; \SEQN{\rvX^{\rLt}}}{\SEQN{\rlambda}} - \CMI{\SEQN{\rvX};
       \SEQN{\rvX^{\rNt_{21}}} }{ \SEQN{\rvX^{\rLt}}, \SEQN{\rlambda} }.
\end{align}
By putting $\SEQN{\rvX^{\rLt}}$ into first mutual information term, we have
\begin{align}
   A & \leq \omega\CMI{\SEQN{\rvX};
       \SEQN{\rvX^{\rLt}},\SEQN{\rvX^{\rMt}}}{ \SEQN{\rlambda}} -\omega \CMI{
            \SEQN{\rvX}; \SEQN{\rvX^{\rLt}}}{\SEQN{\rlambda}} \nonumber \\
      & \qquad - \CMI{\SEQN{\rvX};
       \SEQN{\rvX^{\rNt_{21}}} }{ \SEQN{\rvX^{\rLt}}, \SEQN{\rlambda} } \NONUM \\
      & = \omega\CMI{\SEQN{\rvX};
       \SEQN{\rvX^{\rMt}}}{ \SEQN{\rvX^{\rLt}}, \SEQN{\rlambda}} - \CMI{\SEQN{\rvX};
       \SEQN{\rvX^{\rNt_{21}}} }{ \SEQN{\rvX^{\rLt}}, \SEQN{\rlambda} } \label{eq:partA:t1} \\
    & \leq \sum_{i=1}^n \left[ \omega \CMIR{\SEQ{\rvX}_i;
       \SEQ{\rvX^{\rMt}}_i}{\SEQ{\rvX^{\rLt}}_i, \SEQ{\rlambda}_i,
      \rvD_i} - \CMI{\SEQ{\rvX}_i;
       \SEQ{\rvX^{\rN_{21}}}_i}{\SEQ{\rvX^{\rLt}}_i, \SEQ{\rlambda}_i,
      \rvD_i}  \right] \label{eq:partA:t2}
\end{align}
where~\eqref{eq:partA:t1} is due to chain rule and in~\eqref{eq:partA:t2} we apply Lemma~\ref{lm:marton} with
$\SEQN{\rvX}$ as the channel input and $\SEQN{\rvX^{\rNt_{21}}}$, $\SEQN{\rvX^{\rMt}}$, and $\SEQN{\rvX^{\rLt}}$ are the corresponding
channel outputs. Here,
\begin{align}
\rvD_i := \left(\SEQ{\rvX^{\rMt}}_1^{i+1}, \SEQ{\rvX^{\rNt_{21}}}_{i+1}^n \SEQ{\rvX^{\rLt}}_1^{i-1}, \SEQ{\rvX^{\rLt}}_{i+1}^{n},
  \SEQ{\rlambda}_1^{i-1}, \SEQ{\rlambda}_{i+1}^n\right).\NONUM
\end{align}
Rewrite~\eqref{eq:partA:t2} with entropy, then we have
\begin{align}
   A &\leq \sum_{i=1}^n \left[ \omega \CENTR{\SEQ{\rvX^{\rMt}}_i}{\SEQ{\rvX^{\rLt}}_i, \SEQ{\rlambda}_i,
      \rvD_i} - \CENTR{\SEQ{\rvX^{\rNt_{21}}}_i}{\SEQ{\rvX^{\rLt}}_i, \SEQ{\rlambda}_i,
      \rvD_i}  \right] \label{eq:partA:t3}
\end{align}
In Appendix~\ref{app:partA}, we will show that
\begin{align}
  \CENTR{\SEQ{\rvX^{\rMt}}_i}{\SEQ{\rvX^{\rLt}}_i, \SEQ{\rlambda}_i,
      \rvD_i} &= \sum_{l=1}^q \gamma_1(l) \CENTR{\SEQ{\rX_l}_i}{\SEQ{\rvX_1^{l-1}}_i,
      \rvD_i} \label{eq:A:entropy1} \\
\CENTR{\SEQ{\rvX^{\rNt_{21}}}_i}{\SEQ{\rvX^{\rLt}}_i, \SEQ{\rlambda}_i,
      \rvD_i}  &= \sum_{l=1}^q \alpha_1(l) \CENTR{\SEQ{\rX_l}_i}{\SEQ{\rvX_1^{l-1}}_i,
      \rvD_i} \label{eq:A:entropy2}  
\end{align}
Substitute~\eqref{eq:A:entropy1} and~\eqref{eq:A:entropy2} into~\eqref{eq:partA:t3}, then we have
\begin{align}
  A &\leq \sum_{i=1}^n \sum_{l=1}^q (\omega\gamma_1(l) - \alpha_1(l)) \CENTR{ \SEQ{\rX_l}_i }{ \SEQ{\rvX_1^{l-1}}_i,
       \rvD_i} \NONUM \\
    & \leq n \sum_{l=1}^q  (\omega\gamma_1(l) - \alpha_1(l))^+. \label{eq:A} 
\end{align}

For part $B$, we have two different ways to handle it, which will lead us to bound $\BOUND_{1b}(\omega)$ and bound
$\BOUND_{1c}(\omega, \mu)$, respectively. To obtain $\BOUND_{1b}(\omega)$, we apply Lemma~\ref{lm:mac} with $\rN_1=\rN_{21}$
and $\rN_2=\rN_{11}$ in~\eqref{eq:partB}: 
\begin{align}
  B & \leq n\omega \MEXPP\rN_{21} + \omega\CMI{\SEQN{\rvW}; \SEQN{\rvW^{(\rN_{11}-\rN_{21})^+}} }{ \SEQN{\rvN}} -
    \omega \CMI{\SEQN{\rvW}; \SEQN{\rvW^{\rN_{12}}}}{\SEQN{\rvN}} \NONUM \\
   & \leq n\omega \MEXPP\rN_{21} + \omega \sum_{i=1}^q \left[ \CMI{\SEQ{\rvW}_i;
     \SEQ{\rvW^{(\rN_{11}-\rN_{21})^+}}_i}{\rvD'_i, \SEQ{\rvN}_i }
   -\CMI{\SEQ{\rvW}_i; \SEQ{\rvW^{\rN_{12}}}_i}{\rvD'_i, \SEQ{\rvN}_i}
     \right] \label{eq:B:t1} \\
 & =  n\omega \MEXPP\rN_{21} + \omega \sum_{i=1}^q \left[ \CENT{\SEQ{\rvW^{(\rN_{11}-\rN_{21})^+}}_i}{\rvD'_i,
   \SEQ{\rvN}_i } -\CENT{\SEQ{\rvW^{\rN_{12}}}_i}{\rvD'_i, \SEQ{\rvN}_i}
     \right] \label{eq:B:t2}
\end{align}
where in~\eqref{eq:B:t1}, we apply Lemma~\ref{lm:marton} with $\rvW$ as the channel input and $\rvW^{\rN_{12}}$,
$\rvW^{(\rN_{11}-\rN_{21})^+}$ and a dummy constant as the three channel outputs. Here,
$\rvD'_i := \left(  \SEQ{\rvW^{(\rN_{11}-\rN_{21})^+}}_1^{i-1}, \SEQ{\rvW^{\rN_{12}}}_{i+1}^n,
  \SEQ{\rvN}_1^{i-1}, \SEQ{\rvN}_{i+1}^n\right)$.

Similarly as the proof of~\eqref{eq:A}, we have
\begin{align}
  & \sum_{i=1}^n \left[ \CENT{\SEQ{\rvW^{(\rN_{11}-\rN_{21})^+}}_i}{\rvD'_i,
   \SEQ{\rvN}_i } -\CENT{\SEQ{\rvW^{\rN_{12}}}_i}{\rvD'_i, \SEQ{\rvN}_i} \right] \NONUM \\
 \leq & n\omega\sum_{l=1}^q \left[ \Prob{\rN_{11}-\rN_{21} \geq l} -
      \Prob{\rN_{12} \geq l}\right]^+
\end{align}
Substitute it into~\eqref{eq:B:t2}, we obtain
\begin{align}
  B & \leq n\omega \MEXPP\rN_{21}+ n\omega\sum_{l=1}^q \left[ \Prob{\rN_{11}-\rN_{21} \geq l} -
      \Prob{\rN_{12} \geq l}\right]^+ \label{eq:B1}
\end{align}

With~\eqref{eq:weightCom1}, ~\eqref{eq:A}, and~\eqref{eq:B1} and letting $n \to \infty$, we have
\begin{align}
  R_1 + \omega R_2 &\leq (1-\omega) \MEXPP\rN_{11} + \omega \MEXPP\rN_{12} + \sum_{l=1}^q \left[ \omega \gamma_1(l) -
  \alpha_1(l)\right]^+ + \omega \MEXPP\rN_{21} \nonumber \\
  & \qquad + \omega \sum_{l=1}^q \left[ \Prob{\rN_{11}-\rN_{21} \geq l} -
      \Prob{\rN_{12} \geq l}\right]^+ \NONUM \\
   & =  (1-\omega) \MEXPP\rN_{11} + \omega \MEXPP\rN_{21}  + \sum_{l=1}^q \left[ \omega \gamma_1(l) -
  \alpha_1(l)\right]^+ + \sum_{l=1}^q \max\left( \Prob{\rN_{11}-\rN_{21} \geq l}, \Prob{\rN_{12} \geq l} \right).
\end{align}
which is bound $\BOUND_{1b}(\omega)$.

To show bound $\BOUND_{1c}(\omega, \mu)$, we use the fact that $nR_1 - n \delta_n \leq \CMI{\SEQN{\rvW}; \SEQN{\rvW^{\rN_{11}}}
  }{\SEQN{\rvN}}$. Therefore, with~\eqref{eq:partB} and $\forall \mu \in [0, \omega]$, we have
\begin{align}
  B + n\mu R_1 &- n \mu \delta_n \NONUM \\
   &  \leq  \omega \CMI{\SEQN{\rvW}, \SEQN{\rvX}; \SEQN{\rvW^{\rN_{11}} \oplus
  \rvX^{\rN_{21}}}}{\SEQN{\rvN}} \nonumber \\
    & \quad + \mu \CMI{\SEQN{\rvW}; \SEQN{\rvW^{\rN_{11}}}}{\SEQN{\rvN}} -  \omega
  \CMI{\SEQN{\rvW}; \SEQN{\rvW^{\rN_{12}}}}{\SEQN{\rvN}} \NONUM \\
 & \leq \omega n\MEXP{\max\left( \rN_{11}, \rN_{21}\right)} + \mu \CMI{\SEQN{\rvW}; \SEQN{\rvW^{\rN_{11}}}}{\SEQN{\rvN}} -  \omega
  \CMI{\SEQN{\rvW}; \SEQN{\rvW^{\rN_{12}}}}{\SEQN{\rvN}} \label{eq:B2:t1}
\end{align}
where~\eqref{eq:B2:t1} is due to that the entropy is maximized by setting $\SEQN{\rvX}$ and $\SEQN{\rvW}$ as two independent
i.i.d. sequence with each element of each random vector is an independent $\Ber{1/2}$ random variable.  Now, apply
Lemma~\ref{lm:marton} with $\SEQN{\rvW}$ as the channel inputs and $\SEQN{\rvW^{\rN_{12}}}$, $\SEQN{\rvW^{\rN_{11}}}$, and a dummy constant
as the three channel outputs, then we have 
\begin{align}
    &B + n\mu R_1 - n \mu \delta_n \nonumber \\
  & \leq \omega n\MEXPP\max\left( \rN_{11}, \rN_{21}\right) + \sum_{i=1}^n \mu \CMI{\SEQ{ \rvW}_i;
    \SEQ{\rvW^{\rN_{11}}}_i }{\SEQ{\rvN}_i, \rvD''_i} -  \omega
  \CMI{\SEQ{\rvW}_i; \SEQ{\rvW^{\rN_{12}}}_i}{\SEQ{\rvN}_i, \rvD''_i} \NONUM \\
  & = \omega n\MEXPP\max\left( \rN_{11}, \rN_{21}\right) + \sum_{i=1}^n \mu \CENT{
    \SEQ{\rvW^{\rN_{11}}}_i }{\SEQ{\rvN}_i, \rvD''_i} -  \omega
  \CENT{\SEQ{\rvW^{\rN_{12}}}_i}{\SEQ{\rvN}_i, \rvD''_i}
\end{align}
where $\rvD''_i := \left( \SEQ{\rvW^{\rN_{11}}}_1^{i-1},
  \SEQ{\rvW^{\rN_{12}}}_{i+1}^n,  \SEQ{\rvN}_1^{i-1}, \SEQ{\rvN}_{i+1}^n \right)$.
Similarly as the proof of~\eqref{eq:A} or~\eqref{eq:B2}, we have
\begin{align}
  B + n\mu R_1 - n \mu \delta_n \leq  \omega n\MEXPP\max\left( \rN_{11}, \rN_{21}\right) + n\sum_{i=1}^q \left[ \mu \Prob{\rN_{11} \geq l} -
    \omega\Prob{\rN_{12} \geq l}\right]^+ \label{eq:B2}
\end{align}

Combining \eqref{eq:weightCom1}, \eqref{eq:A}, and \eqref{eq:B2} and
letting $n \to \infty$, we have
\begin{align}
  (1+\mu)R_1 + \omega R_2  & \leq (1-\omega) \MEXPP\rN_{11} + \omega \MEXPP\rN_{12} + \sum_{l=1}^q \left[ \omega \gamma_1(l) -
  \alpha_1(l)\right]^+ \NONUM \\
   & + \omega \MEXP{\max\left( \rN_{11}, \rN_{21}\right)} + \sum_{l=1}^q \left[\mu \Prob{\rN_{11} \geq l} -
    \omega\Prob{\rN_{12} \geq l}\right]^+ \NONUM \\
  & = \MEXPP\rN_{11} + \omega\MEXP{ \rN_{21} - \rN_{11}}^+ + \sum_{l=1}^q \left[ \omega \gamma_1(l) -
  \alpha_1(l)\right]^++  \sum_{l=1}^q \max\left(  \mu \Prob{\rN_{11} \geq l},
    \omega\Prob{\rN_{12} \geq l} \right)
\end{align}
Thus, bound $\BOUND_{1c}(\omega, \mu)$ holds for $\forall \omega \in [0,1]$ and $\forall \mu \in [0, \mu]$.

\section{Concluding Remarks}\label{sec:conclude}
In this paper, an outer bound for general two-user layered erasure interference channel is derived. It is tight in
several important cases but remaining open in others. As we pointed out above, the major roadblock to fully close this problem
is the moderate interference case for $q=1$. For that particular case, the best known inner bound derived in~\cite{Vahid2015IC} does not meet with
our new outer bound either. Future work will extend the study here to Gaussian fading interference channels.

\appendices

\section{Proof of~\eqref{eq:det} via Theorem~\ref{thm:1}}\label{app:det}
Following summary in Table~\ref{tab:1}, let $\omega = 0$ in $\BOUND_{1a}(\omega)$ and
$\BOUND_{2a}(\omega)$, respectively, then we obtain~\eqref{eq:det:a}. Let $\omega = 1$ for bound $\BOUND_{1a}(\omega)$, we
have
\begin{align}
  R_1 + R_2  &\leq n_{11} + (n_{21}-n_{11})^+ + \sum_{l=1}^q \left[ \Prob{n_{22}\geq l} - \Prob{n_{21} \geq l} \right]^+
               \NONUM \\
      & = \max(n_{11}, n_{21})  + (n_{22}-n_{21})^+  \NONUM
\end{align}
which recovers~\eqref{eq:det:b}. By symmetry, bound $\BOUND_{2a}(1)$ can indicates~\eqref{eq:det:c}. Next, consider
bound $\BOUND_{1b}(1)$, we obtain
\begin{align}
  R_1 + R_2 & \leq n_{21} + \sum_{l=1}^q \left[\Prob{n_{22} - n_{12} \geq l} - \Prob{n_{21} \geq l} \right]^+ +
  \sum_{l=1}^q \max\big(\Prob{n_{11} - n_{21} \geq l}, \Prob{n_{12}\geq l}\big)  \NONUM \\
    & = n_{21} + \big((n_{22}-n_{12})^+ - n_{21}\big)^+ + \max\big((n_{11}-n_{21})^+, n_{12}\big) \NONUM \\
    & = \max\big((n_{11}-n_{21})^+, n_{12}\big)  +  \max((n_{22}-n_{12})^+, n_{21}) \NONUM 
\end{align}
which coincides with~\eqref{eq:det:d}. By symmetry, we can conclude that bound $\BOUND_{2b}(1)$ can also
recover~\eqref{eq:det:d}. To obtain~\eqref{eq:det:e}, we consider $\BOUND_{1c}(1,1)$, \IE, $\omega = \mu = 1$:
\begin{align}
  2R_1 + R_2 & \leq n_{11} + (n_{11}-n_{21})^+ + \sum_{l=1}^q \left[\Prob{n_{22} - n_{12} \geq l} - \Prob{n_{21} \geq l}
  \right]^+ \nonumber \\
  & \qquad + \sum_{l=1}^q \max(\Prob{n_{11} \geq l}, \Prob{n_{12} \geq l}) \NONUM \\
 & = \max(n_{11}, n_{21}) + (n_{22}-n_{12})^+ - n_{21})^+ + \max(n_{11}, n_{12}) \NONUM \\
& =  (n_{11}-n_{21})^+ + n_{21} +  \big((n_{22}-n_{12})^+ - n_{21}\big)^+ + \max(n_{11}, n_{12}) \NONUM \\
 & = (n_{11}-n_{21})^+ + \max\big((n_{22}-n_{12})^+, n_{21}\big)  + \max(n_{11}, n_{12}) 
\end{align}
which is same as~\eqref{eq:det:e}. By symmetry, \eqref{eq:det:f} can be obtained by bound $\BOUND_{2c}(1,1)$. This
completes our proof.

\section{Proof of~\eqref{eq:A:entropy1} and~\eqref{eq:A:entropy2}}\label{app:partA}
For~\eqref{eq:A:entropy1},
\begin{align}
  &\CENTR{\SEQ{\rvX^{\rMt}}_i}{\SEQ{\rvX^{\rLt}}_i, \SEQ{\rlambda}_i, \rvD_i} \NONUM \\
 = & \sum_{\sLt=1}^q \sum_{\sMt=1}^q \Prob{\rMt = \sMt, \rLt = \sLt} \CENTR{\SEQ{\rvX^{\sMt}}_i}{\SEQ{\rvX^{\sLt}}_i, \rvD_i}
     \NONUM \\
= & \sum_{\sLt=1}^q \sum_{\sMt=1}^q \Prob{\rMt = \sMt, \rLt = \sLt}  \sum_{l=1}^q \IND{\sLt < l \leq \sMt}
    \CENTR{\SEQ{\rX_l}_i}{\SEQ{\rvX^{l-1}}_i, \rvD_i} \label{eq:A:entropy1:t1} \\
 = &\sum_{l=1}^q \Prob{\rLt < l \leq \rMt}  \CENTR{\SEQ{\rX_l}_i}{\SEQ{\rvX^{l-1}}_i, \rvD_i} 
\end{align}
where we apply chain rule in~\eqref{eq:A:entropy1:t1}. Furthermore, since $\rLt$ and $\rMt$ are coupled
as~\eqref{eq:alignment}, we have
\begin{align}
  \Prob{\rLt < l \leq \rMt}  & = \Prob{\rMt \geq l } - \Prob{\rMt \geq l, \rLt \geq l} \NONUM \\
                                          & = \Prob{\rMt \geq l} - \min\left( \Prob{\rMt \geq l }, \Prob{\rLt \geq l }
                                            \right) \NONUM \\
                                          & = \left[ \Prob{\rMt \geq l } - \Prob{\rLt \geq l } \right]^+ \NONUM \\
                                          & = \big[ \Prob{ \rN_{22} - \rN_{12} \geq l } - \Prob{\rN_{21} - \rN_{11} \geq
                                            l } \big]^+ \label{eq:A:entropy1:t2}
\end{align}
where~\eqref{eq:A:entropy1:t2} is due the fact that $\rMt \sim (\rN_{22} - \rN_{12})^+$, $\rLt \sim (\rN_{21} - \rN_{11})^+$,
and $l > 0$.

We can prove~\eqref{eq:A:entropy2} similarly:
\begin{align}
  &\CENTR{\SEQ{\rvX^{\rNt_{21}}}_i}{\SEQ{\rvX^{\rLt}}_i, \SEQ{\rlambda}_i,
      \rvD_i}  \NONUM \\
= &\sum_{\sLt=1}^q \sum_{\sNt_{21}=1}^q  \Prob{\rNt_{21} = \sNt_{21}, \rLt = \sLt} \CENTR{\SEQ{\rvX^{\sNt_{21}}}_i}{\SEQ{\rvX^{\sLt}}_i,
      \rvD_i}  \NONUM \\
= &\sum_{\sLt=1}^q \sum_{\sNt_{21}=1}^q  \Prob{\rNt_{21} = \sNt_{21}, \rLt = \sLt} \sum_{l=1}^q \IND{\sLt < l
  \leq \sNt_{21}} \CENTR{\SEQ{\rX_l}_i}{\SEQ{\rvX^{l-1}}_i,
      \rvD_i}  \NONUM \\
= & \sum_{l=1}^q \Prob{ \rLt < l \leq \rNt_{21}} \CENTR{\SEQ{\rX_l}_i}{\SEQ{\rvX^{l-1}}_i,
      \rvD_i} 
\end{align}
Furthermore, note that under coupling~\eqref{eq:alignment}, we have $\rNt_{21} \geq \rLt$. Therefore, 
\begin{align}
  \Prob{ \rLt < l \leq \rNt_{21}} & = \Prob{\rNt_{21} \geq l } - \Prob{\rLt \geq l, \rNt_{21} \geq l} \NONUM \\
 & =  \Prob{\rNt_{21} \geq l} - \Prob{\rLt \geq l} \NONUM \\
 & =  \Prob{\rN_{21} \geq l} - \Prob{\rN_{21} - \rN_{11} \geq l}
\end{align}
where in the last equality, we use the fact that $\rN_{21} \sim \rNt_{21}$, $\rLt \sim (\rN_{21} -\rN_{11})^+$, and $l
> 0$.

\bibliographystyle{IEEEtran}
\bibliography{IEEEabrv,zybibset,zybib_IC}
\end{document}